\documentclass{Interspeech2024}

\usepackage{bbm}

\DeclareMathOperator*{\argmin}{argmin\,}

\interspeechcameraready

\title{Towards a Universal Method for Meaningful Signal Detection}

\name{Louis}{Mahon}

\address{
  University of Edinburgh, UK
  }
\email{lmahon@ed.ac.uk}


\begin{document}

\maketitle

\begin{abstract}
    It is known that human speech and certain animal vocalizations can convey meaningful content because we can decipher the content that a given utterance does convey. This paper explores an alternative approach to determining whether a signal is meaningful, one that analyzes only the signal itself and is independent of what the conveyed meaning might be. We devise a method that takes a waveform as input and outputs a score indicating its degree of `meaningfulness`. We cluster contiguous portions of the input to minimize the total description length, and then take the length of the code of the assigned cluster labels as meaningfulness score. We evaluate our method empirically, against several baselines, and show that it is the only one to give a high score to human speech in various languages and with various speakers, a moderate score to animal vocalizations from birds and orcas, and a low score to ambient noise from various sources.
\end{abstract}

\section{Introduction} \label{sec:intro}
Humans are highly proficient at auditory pattern recognition, at least for certain subsets of audio signals, such as those of human speech. We are are able to interpret a wide variety of meanings across a wide variety of waveforms. Similarly, when we regard an animal vocalizations as meaningful, it is by finding the behaviour or information that we believe it signals. However, even independently of interpreting what a given signal means, humans also possess some ability to detect whether a signal is meaningful. For example, if one hears speech in a language they do not understand, or certain animal vocalizations, they may still get a sense that this is the sort of signal that could convey meaning. We do not feel the same if we hear other types of signal, such as ambient noise or white noise on the radio. Some signals, and data more generally, exhibit a systematic structure that suggests the potential to convey a meaning. Humans have a degree of intuition for recognizing this sort of structure. The goal of this paper is to make first steps towards articulating what it is, and how we might measure it automatically.

There exist several classic approaches to measuring complexity. Kolmogorov complexity (KC) takes the length of the shortest program that generates the given data. It is uncomputable, but can be approximated by file-compression ratio. The minimum description length principle \cite{grunwald2007minimum} (MDL) principle is an information-theoretic approach to the same idea that the one should search for the shortest description of the data. Mostly, MDL is concerned with model selection, though has also been used for other applications such as a parameter-free algorithm for optimal segmentation \cite{mahon-lapata-2024-modular}. \cite{mahon2024minimum} has used MDL for measuring complexity. Entropy is closely related to MDL, and is often used as a measure of complexity for data of various sorts \cite{ribeiro2017characterizing, costa2002multiscale}. One problem that arises with these methods, and their many derivatives, is that they give a low score to simple, highly regular data, and a high score to random, noisy data, with the data we consider most meaningful, such as human speech, falling somewhere between the two. Thus, we can say neither that a high score nor a low score indicates the data is meaningful. This fundamental issue has been identified by various authors \cite{mahon2024minimum, ay2010effective,vitanyi2006meaningful,gell1996information,koppel1987complexity}.

We follow the MDL principle for our method, but make a novel division of the description into a `meaningful' and `meaningless' portion. This is similar, on a high level, to some theoretical work to divide KC into meaningful information and noise, using, e.g., `sophistication' \cite{koppel1987complexity,vitanyi2006meaningful} or `effective complexity' \cite{ay2010effective,gell1996information}. When selecting the shortest description of the data, we include both the meaningful and the meaningless portion, but, after making this selection, we ignore the meaningless portion and take only the length of the meaningful portion as contributing to the complexity. Put semi-formally, if $d$ is a description of data $X$, and $m(d)$ is the meaningful portion of $d$, then our proposed meaningfulness score is given by
$|m(d^*)|$, where $d^* = \argmin_d |d(X)|$. Note that the meaningless portion still plays an important role, in selecting the optimal description. 

Of course, the meaning of a signal depends not just on the structure of the signal itself, but also on the surrounding social context. What we investigate here might more accurately be called ``potential [for a signal] to be meaningful given the right context''. With this caveat in mind, for the sake of concision, we refer to this just as ``meaningfulness''.

The contributions of this paper are the following:
\begin{itemize}
    \item the articulation of the problem of characterizing meaningfulness and why existing methods are inadequate;
    \item the description of a method that avoids these shortcomings and is able to give a high score to data we know to be meaningful, and a low score to random or simple uniform data;
    \item the empirical evaluation of this method, compared against several baselines, on a variety of signal types.
\end{itemize}

In the remainder of this paper, Section \ref{sec:related-work} gives an overview of related work, Section \ref{sec:method} describes our method in detail, Section \ref{sec:experimental-eval} presents the empirical evaluation, and Section \ref{sec:conclusion} outlines future work and summarizes.

\section{Related Work} \label{sec:related-work}
The problem of measuring the complexity of data has mostly been studied in the visual domain, that is, in measuring the complexity of an image. Some methods use file compression ratio, GIF and TIFF in \cite{marin2013examining} and JPEG in \cite{machado2015computerized}, claiming that a lower ratio means high complexity. Others have used the gradient of pixel intensities \cite{redies2012phog} or fractal dimension \cite{sun2006fractal}. In \cite{mahon2024minimum}, it was shown that these approaches fail to distinguish meaningful complexity from noise, and give a maximum score to white noise images. Instead, they propose to cluster patches of the image, using the MDL principle to select outliers and the number of clusters, then take the entropy of cluster indices that appear inside each patch. Our method is inspired by that of \cite{mahon2024minimum}, but differs in two respects. Firstly, it applies to the one-dimensional case of signal processing, rather than the two-dimensional case of images, which means the patch-based recursive procedure used in \cite{mahon2024minimum} cannot apply. Secondly, we omit the complicated calculation of entropy from regions of cluster indices that \cite{mahon2024minimum} uses, and instead invoke the distinction between the meaningful and meaningless portions of the description.

In the signal processing domain, several works have proposed to measure complexity using some variant of entropy, such as evaluating on multiple timescales \cite{costa2002multiscale}, using Tsallis q entropy \cite{ribeiro2017characterizing},  or replacing sections of the waveform with discrete symbols \cite{liu2016increment}.
Unsupervised analysis of speech and animal vocalizations has mostly focused either on combining clustering and deep learning for feature extraction \cite{mahon2023efficient,zeghidour2021wavesplit,mahon2024hard}, or on acoustic unit discovery. Some works have applied the deep learning plus clustering approach specifically to animal vocalizations, such as distinguishing different species' vocalizations \cite{guerrero2023acoustic}, or distinguishing call types within a single population of orcas \cite{bergler2019deep}. In terms of calculating time series complexity, one method that has been used by several works \cite{ali2018innovative, yazdi2021complexity} is to take the fractal dimension, as calculated by the Katz method \cite{katz1988fractals}. In Section \ref{sec:experimental-eval}, we show empirically that our method is better able to distinguish different signal types than the Katz fractal dimension, as well as entropy and compression-ratio.


\section{Method} \label{sec:method}
We assume we are presented with some set of data points, and want to assign it a meaningfulness score. 
We cluster the data and represent each point by first specifying its assigned cluster, and then specifying where it falls in that cluster`s distribution. The former, which we take as the meaningful portion, comprises an index from $1, \dots, K$. The latter, the meaningless portion, could admit many different coding schemes, but by the Kraft-McMillan inequality, we know that, under the optimal coding scheme, the length will be bounded by, and close to $- \log p(x)$, where $p(x)$ is the probabilty under the cluster`s distribution. 

Alternatively, a data point can be specified directly, independently of its assigned cluster. For example, if the data consists of 64-dimensional vectors of 32-bit floats, it can be specified directly with 64 $\times$ 32 = 2048 bits. In this case, we regard the entire description for that data point as meaningless, as it is far away from its cluster centroid, suggesting it does not fit into a coherent pattern alongside the rest of the data points. 

For a given clustering partition and given data point, we choose either the cluster-based description, or the cluster-independent description, whichever is smaller. We also take into account the number of bits needed to directly specify the clustering model itself, such as the cluster centroids (the exact parameters depend on the clustering method used). This imposes a slight additional cost on having a larger number of clusters. The total description length under a given partition is the description length of the model plus the sum of the lengths of the description of each data point under that partition, and the partition is selected to minimize this overall description length. Once the optimal partition has been found, we add together the length of the meaningful portion of the description for each data point, which amounts to taking the Shannon information content of the cluster labels assigned to all data points that use the cluster-based description, plus the description length of the model itself. The resulting sum is the final meaningfulness score.

\begin{table*}[]
    \centering
    \caption{Comparison of the mean (with std in brackets) scores given by our method for each type of signal, compared with four baseline methods. Only ours gives speech a very high score, animal vocalizations a high score, and other sounds a low score.}
    \label{tab:main-results}
\resizebox{0.9\textwidth}{!}{
\begin{tabular}{llllll}
\toprule
 & ours & katz & ent & zl comp ratio & wav comp ratio \\
\midrule
walla & 61.3 (1.78) & 12.9 (1.18) & 100.0 (0.01) & 21.7 (1.51) & 14.8 (0.63) \\
tuning-fork & 43.6 (7.93) & 48.4 (8.31) & 86.1 (9.79) & 21.3 (4.80) & 15.0 (3.86) \\
birdsong & 72.8 (4.11) & 5.1 (0.75) & 100.0 (0.01) & 10.8 (1.10) & 22.6 (0.84) \\
birdsong-background & 18.5 (6.84) & 0.0 (0.01) & 99.9 (0.03) & 1.7 (0.48) & 28.3 (0.07) \\
orcavoc & 75.1 (2.95) & 35.6 (8.15) & 100.0 (0.01) & 19.1 (4.40) & 10.8 (2.36) \\
orcavoc-background & 29.0 (5.15) & 6.2 (1.19) & 100.0 (0.01) & 17.5 (7.36) & 21.3 (0.70) \\
irish-m-speech & 83.8 (2.14) & 10.0 (1.68) & 100.0 (0.01) & 40.7 (4.45) & 83.0 (3.36) \\
irish-f-speech & 83.1 (2.84) & 12.5 (2.69) & 98.0 (1.83) & 35.2 (2.84) & 54.6 (10.00) \\
german-m-speech & 84.1 (3.81) & 12.4 (1.55) & 100.0 (0.01) & 67.6 (7.25) & 29.5 (2.20) \\
german-f-speech & 88.3 (1.89) & 17.4 (1.18) & 100.0 (0.01) & 35.1 (4.87) & 50.4 (1.40) \\
english-m-speech & 84.6 (2.43) & 16.4 (2.31) & 100.0 (0.01) & 35.5 (3.51) & 20.2 (1.36) \\
english-f-speech & 85.0 (2.71) & 25.7 (4.53) & 100.0 (0.01) & 37.6 (5.02) & 21.4 (2.26) \\
rain & 2.1 (0.24) & 25.4 (0.78) & 100.0 (0.01) & 7.4 (0.24) & 4.7 (0.30) \\
\bottomrule
\end{tabular}

}
\end{table*}

\subsection{Formal Description}
Let $X \subset \mathbb{R}^m, X = x_1,\dots, x_n$ be the input data. Let $c$ be the numerical precision, e.g. $c=32$ in the case of representing real numbers with 32-bit floats. Let $p(x;\mu, \Sigma)$ be the multivariate normal probability of data point $x \in \mathbb{R}^m$ given cluster centroid $\mu \in \mathbb{R}^m$ and diagonal covariance matrix $\Sigma \in \mathbb{R}^m$ (we consider only diagonal covariances to speed up search). Let $g$ be the function that takes as input a partition function $f:\mathbb{R}^m \rightarrow \{0, \dots, n-1\}$, and a data point $x \in X$, and returns the centroid of $x$ under partition $f$. That is $g(f, x) = \frac{1}{|C|} \sum_{y \in C} y$, where $C = \{y \in X | f(y)=f(x)\}$. Let $h$ be the analogous function that returns the diagonal of the covariance matrix of the cluster of point $x$ under $f$, and let $q(x,f) = p(x; g(x, f), h(x,f))$. Let $l(i, f)$ give the number of points assigned to the $i$th cluster under the partition $f$. Then the fit clustering model is given by
\begin{gather} \label{eq:optimal-partition}
    f^* = \argmin_{f:\mathbb{R}^m \rightarrow \{1, \dots, n\}} \sum_{i=0}^n \min(cm, -\log{q(x_i, f)}) \\
    + \sum_{i=1}^n \log{\frac{n}{l(f(i), f)}}\mathbbm{1}(-\log{q(x_i, f)} < cm)\,,
\end{gather}

where the last term uses the indicator function $\mathbbm{1}$ to select only those points whose cluster-based description cost is less than their cluster-independent description cost, and the sum represents the Shannon information content of the cluster labels of those points. This selection of the partition that allows the shortest overall description of the data follows the MDL principle. Then, the complexity score is given by
\begin{gather} \label{eq:comp-score}
\sum_{i=1}^n \log{\frac{n}{l(i, f^*)}}\mathbbm{1}(-\log{q(x_i, f)} < cm) \\
+ 2cm \sum_{i=1}^n \mathbbm{1}(l(i, f^*) > 0)\,,
\end{gather}
where the second sum is for the description of the model itself: the mean and a covariance diagonal vector of each cluster. 

\subsection{Implementation Details}
We use a Gaussian mixture model (GMM) for clustering
The GMMs are initialized with k-means, use diagonal covariance matrices, and are optimized with the usual expectation-maximization algorithm, with tolerance $1e-3$, capped at $100$ iterations. They are fit 10 times with random initializations and the we select the one with the highest data probability. In order to optimize the number of clusters, we fit a separate GMM with $K$ clusters for $K = 1, \dots, 8$, and keep the partition from the one with the lowest cost, as given by \eqref{eq:optimal-partition}. We take a single waveform as input, and form a spectrogram (window size = fft size = 30, overlap = 3). The fft for each window (i.e. column of the spectgrom) is then taken as a data point, and so we treat the single waveform as a dataset on which to run our method.

We then repeat our method twice more, where instead of taking each individual segment as a separate data point, we take contiguous chunks of several segments, 2 on the first repeat and 4 on the second. At levels two and three, the vector of each data point is not the frequency spectrum, but rather the multiset of cluster indices, from the previous level, found in chunk centred at that point, e.g. if the chunk contained two points that were assigned to the first cluster, none to the second and one to the third, the vector would be $[2,0,1]$. This is to allow the method the potential to capture higher-level compositional structure, such as found in language.

\section{Experimental Evaluation} \label{sec:experimental-eval}
In this section, we show the scores from applying our method to different types of signals. Typical machine learning methods target only a particular domain and aim to distinguish different classes within that domain, e.g. distinguish between different phonemes or different speakers from human speech in a given language. Ours, in contrast, is a general method, designed to apply to any waveform with no restrictions on domain. Therefore, we evaluate it on various different types of signal, and report the average score for each signal type. Our method operates separately on each waveform, and requires no training data.

\subsection{Datasets}
The signal types we consider are birdsong, orca vocalizations, the background noise in these recordings, human speech in English, Irish and German, and two types of ambient noise: rainfall and muffled overlapping human conversations, a.k.a. `walla'. We also consider recordings of tuning forks, which are physical musical devices designed to give a pure tone when struck. Walla and rainfall are public recordings from \url{https://www.soundjay.com}. The orca vocalizations we use comprise discrete calls only and are taken from public domain recordings by the National Park Service (NPS), available at \url{https://archive.org/details/KillerWhaleorcinusOrcaSoundsVocalizations}. Birdsong recordings are from the Powdermill acoustic dataset recorded in the Powdermill Nature Reserve, PA, and comprise Black-throated Green Warbler (BTNW), Ovenbird (OVEN), and Eastern Towhee (EATO). Speech recordings are taken from the common voice project, with one male and one female speaker of each language. 
We take time intervals of 1s at 44100 Hz, for all signal types. To show that our method is not simply responding to vocalization having greater amplitude than background noise, we normalize all waveforms to the same mean amplitude. 
For all datasets, we randomly pick 10 utterances per class, and manually select sections with vocalizations (ot without vocalizations, in the `without` setting). 

\subsection{Comparison Methods}
Several existing methods purport to measure data complexity. Some authors have proposed variations of entropy for this purpose \cite{hughes1993analysis, mateos2018measures}. Here we compare our method to a baseline that takes the Shannon entropy of the spectrogram of the input signal. Another approach is to use the file compression ratio, This has been used in opposite senses, \cite{nagaraj2013new} claim that a more `complex' signal will be less compressible, whereas \cite{rosete2018using} claim that a `communication' signal will be \emph{more} compressible. We evaluate the file-compression ratio, both of the waveform and of the spectrogram, in order to see if there is a pattern in either direction. We use FLAC compression for audio and Zempel-Liv on the image of the spectrogram. These are deliberately selected to be lossless because we are interested in a universal measure of meaningfulness so do not want to introduce assumptions about what parts of the information can be discarded, as e.g. in MP3 compression which invokes human audio perception and psychoacoustics. Finally, we also compare to the Katz fractal dimension, as used in \cite{ali2018innovative}.

\subsection{Main Results}

Table \ref{tab:main-results} shows the scores produced by our method, and the four comparison methods, for each signal type. Our method gives the highest score to the three human languages, English, Irish and German, followed by bird and orca vocalizations, which both get a similar score, and the lowest score to all the sounds that are not vocalizations: the background noise of the birds and orcas, tuning forks, rainfall and muffled human conversation. 

This aligns with our existing understanding of the amount of information conveyed by each of these signal types. We know human speech is highly meaningful, and it is interesting that our method gives a very similar score to the three different languages, and six different speakers. Our scoring is consistent with the general principle that all human languages are roughly equally efficient at conveying meaning \cite{liversedge2016universality, hawkins2014cross, rubio2021speakers}. 

We can be relatively sure that tuning forks, rainfall, and ambient noise are low in meaningfulness. Animal vocalizations are less well understood. They is strong evidence that orca vocalization \cite{sandholm2023orcas, saulitis2005vocal} convey meaning but whether it is as rich as human speech remains an open question. It is therefore correct of our method to give them a higher score than the meaningless baselines, and not unreasonable to be lower than human speech. We also note that the available recordings of vocalizations may not reflect the full meaning being conveyed. One second of human speech contains several phonemes and so spans some diversity of its sound inventory, but the right scale at which to interprete orca vocalizations is not clear.
In the absence of the field`s understanding of the semantic units in orca vocalization, we should regard the figures from Table \ref{tab:main-results} as a lower bound on their amount of meaningful content. 

The comparison methods do not produce the same distinction of the different human speech signals as highly meaningful. The `katz' and `entropy' methods completely fail to show a systematic distinction, with `entropy' assigning all methods essentially the same score, and `katz' giving scores that appear random. The file compression ratio methods, especially, Zempel-Liv, fare better, and generally give speech a high score. This supports the claim of \cite{rosete2018using} that meaningful signals are more compressible, vs \cite{nagaraj2013new} who claimed they were less compressible. However, Zempel-Liv compression ratio is also high for uniform simple inputs, and we can see this in it giving a higher score to the tuning fork than to the animal vocalizations. Compression ratio is an approximation to inverse Kolmogorov complexity, which we have argued is the wrong theoretical approach to quantifying meaningfulness. The overly high score given to simple signals such as the tuning fork is a manifestation of this.

\subsection{Ablation Studies}
Table \ref{tab:abl-results} shows the results of removing two main parts of our method. In `no-mdl', we do not use the minimum description length to select the number of clusters $K$, instead we fix $K=5$ for all inputs and all levels. In this setting, the scores are similar for all signal types. Aside from English speech, which is slightly higher than the others, it fails to distinguish more from less meaningful signals. In `just-one-level', we omit the recursive clustering procedure described in Section \ref{sec:method}, and only the score from the first level. This performs similarly to the full method, still managing to roughly group the signal types into human speech as one group, animal vocalizations as another group, and background/meaningless noise as a third. This shows that the higher levels are only minimally utilized. We expect that future extensions, perhaps with segment lengths that depend on the input, will show a benefit from the higher layers.




\begin{table}[]
    \centering
    \caption{Ablation studies, removing the recursive clustering procedure (`just one level') and the MDL-based selection of outliers and the number of clusters (`no mdl').}
    \label{tab:abl-results}
\resizebox{\columnwidth}{!}{
\begin{tabular}{llll}
\toprule
 & ours & just one level & no mdl \\
\midrule
walla & 62.5 (2.05) & 29.6 (2.02) & 73.1 (1.74) \\
tuning-fork & 44.9 (8.11) & 27.0 (5.47) & 67.5 (6.60) \\
birdsong & 75.6 (3.76) & 42.8 (4.21) & 77.6 (2.18) \\
birdsong-background & 19.4 (6.91) & 7.8 (3.17) & 68.3 (2.72) \\
orcavoc & 74.9 (2.76) & 46.4 (3.10) & 60.8 (6.69) \\
orcavoc-background & 29.7 (5.17) & 12.5 (2.36) & 47.5 (6.65) \\
irish-m-speech & 84.8 (2.26) & 63.8 (3.77) & 81.4 (1.80) \\
irish-f-speech & 87.9 (2.36) & 65.3 (4.04) & 80.3 (3.21) \\
german-m-speech & 89.8 (2.31) & 70.2 (4.45) & 76.5 (4.99) \\
german-f-speech & 89.3 (1.54) & 68.6 (2.95) & 82.2 (1.72) \\
english-m-speech & 86.9 (2.47) & 63.8 (4.49) & 82.1 (4.62) \\
english-f-speech & 87.7 (2.72) & 70.3 (4.59) & 66.5 (7.74) \\
rain & 2.6 (0.31) & 0.2 (0.01) & 70.9 (2.99) \\
\bottomrule
\end{tabular}
}

\end{table}

\begin{figure}
    \centering
    \includegraphics[width=1.00\columnwidth, trim={1cm 0 0.4cm 0}]{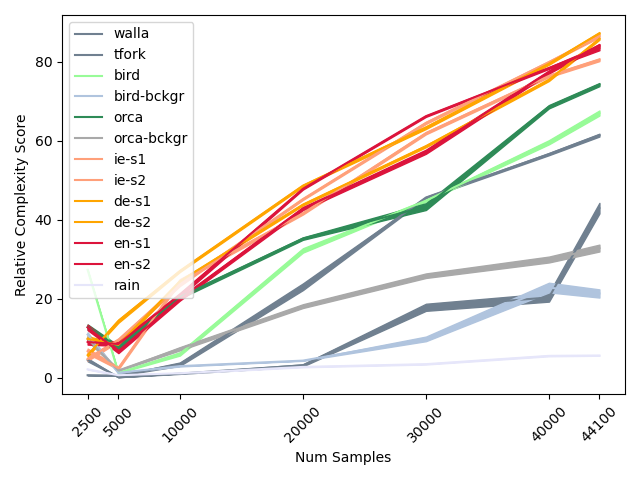}
    \caption{Scores of our method for each signal type, as a function of the number of samples (sample rate 44100 Hz). Colours are rougly grouped by signal type, red-orange for speech, green for animal vocalizations and blue-grey for others. For each language, the first speaker (`-s1') is male, and the second female.}
    \label{fig:nsamples}
\end{figure}

\subsection{Signal Length} \label{subsec:signal-length}
In order to show how the score of our method depends on the size of the input, Figure \ref{fig:nsamples} plots the scores for each signal type as a function of the number of samples. The sample rate is held constant, so fewer samples equats to spanning a smaller time window. The rightmost point for each line corresponds to the main results presented in Table \ref{tab:main-results}, of 1s at 44100 Hz. 

For very low numbers of samples, the method gives all inputs a similar score. However, when the number of samples increases to roughly 20000, it is largely able to distinguish speech, animal vocalizations and ambient noises from each other. This shows that, with as little as 0.5s of audio, our method can assign reasonable meaningfulness scores to the signal types presented.

\section{Conclusion} \label{sec:conclusion}
This paper presented a novel method for quantifying meaningfulness of data, and applied this to method in the domain of signal processing. The meaningfulness score was very high for human speech, across six speakers and three languages, high for birdsong and orca vocalizations, and low for various other signal types, including ambient noise and  the pure tone of tuning forks. The method involves clustering segments of the input signal, so as to minimize the total description length of the data under that clustering, and then taking the Shannon entropy of the cluster labels. To our knowledge, this is the first metric to give a low score to both simple uniform signals, and random noisy signals, while still giving a high score to sounds we know to be meaningful such as human speech. Future work includes, allowing variable length sound segments to adapt to different timescales in vocalization, and testing on a wider variety of animals, speakers, languages and other sound sources.

\bibliographystyle{IEEEtran}
\bibliography{mybib}

\end{document}